\documentclass{an}
\usepackage{graphicx}
\usepackage{times}
\usepackage{fancyhdr}
\newcommand{\gtrsim}{\ ^{\displaystyle >}_{\displaystyle \sim}\ }
\newcommand{\lesssim}{\ ^{\displaystyle <}_{\displaystyle \sim}\ }

\begin{document}

\title{The Generation of Magnetic Fields and X-ray Observations}

\author{Yutaka Fujita\inst{1}
\and  Tsunehiko
N. Kato\inst{2}}
\institute{
Department of Earth and Space Science, 
Graduate School of Science,
Osaka University, 
Toyonaka, Osaka 560-0043, Japan
\and 
National Astronomical Observatory, Osawa 2-21-1,
Mitaka, Tokyo 181-8588, Japan}

\date{Received; accepted; published online}

\abstract{We show that strong magnetic fields can be generated at 
shock waves
associated with formation of galaxies or clusters of galaxies by the
Weibel instability, an instability in collisionless plasmas.  The
estimated strength of the magnetic field generated through this
mechanism is close to the order of values observed in galaxies or
clusters of galaxies at present, which indicates that strong
amplification of magnetic fields after formation of galaxies or clusters
of galaxies is not required.  This mechanism could have worked even at a
redshift of $\sim 10$, and therefore the generated magnetic fields may
have affected the formation of stars at the early universe. This model
will be confirmed by future observations of nearby clusters of
galaxies.  In this context, we also present the Japanese X-ray missions.
}

\correspondence{fujita@vega.ess.sci.osaka-u.ac.jp}

\maketitle

\section{Introduction}

The question of the origin of galactic magnetic fields is one of the
most challenging problems in modern astrophysics. It is generally
assumed that magnetic fields in spiral galaxies are amplified and
maintained by a dynamo through rotation of the galaxies (Widrow 2002).
The dynamo requires seed fields to be amplified. However, observations
of microgauss fields in galaxies at moderate redshifts strongly
constrain the lower boundary of the seed fields (Athreya et
al. 1998). Moreover, magnetic fields are also observed in elliptical
galaxies and galaxy clusters  where rotation, and thus dynamo mechanism
cannot play a central role (Clarke, Kronberg \& B{\" o}hringer 2001;
Widrow 2002; Vall{\' e}e 2004).

On the other hand, the Weibel instability can provide another mechanism
to generate strong magnetic fields (Weibel 1959; Fried 1959).  This
instability is driven in a collisionless plasma, or a tenuous ionized
gas, by the anisotropy of the particle velocity distribution function
(PDF) of the plasma. When the PDF is anisotropic, currents and then
magnetic fields are generated in the plasma so that the plasma particles
are deflected and the PDF becomes isotropic (Medvedev \& Loeb
1999). Through the instability, the free energy attributed to the PDF
anisotropy is transferred to magnetic field energy. This instability
does not need seed magnetic fields. It can be saturated only by
nonlinear effects, and thus the magnetic fields can be amplified to very
high values. This instability has been observed directly in recent laser
experiments (Wei et al. 2002). In astrophysical plasmas, the instability
is expected to develop at shocks or at steep temperature gradients,
where the PDF is anisotropic. Examples of the sites are pulsar winds,
shocks produced by gamma-ray bursts, jets from active galactic nuclei
(AGNs), cosmological shocks, and cold fronts (contact discontinuities
between cold and hot gas) in clusters of galaxies (Medvedev \& Loeb
1999; Kazimura et al. 1998; Nishikawa et al. 2003; Schlickeiser \&
Shukla 2003; Okabe \& Hattori 2003). Although the instability was found
in 1959, its nonlinear nature had prevented us from understanding its
long-term evolution. Recently, however, as computer power increases,
detailed particle simulations of plasmas have been initiated and they
have revealed the evolution of magnetic fields even after saturation of
the instability (Silva et al. 2003; Medvedev et al. 2005). Based on
these results, we consider the generation of magnetic fields at galaxy
and cluster-scale shocks through the Weibel instability at the formation
of galaxies (both ellipticals and spirals) and clusters. We use the
cosmological parameters of $\Omega_0=0.3$, $\lambda=0.7$, the Hubble
constant of $H_0=70\rm\: km\: s^{-1}\: Mpc^{-1}$, and $\sigma_8=0.9$.

\section{Models}
\subsection{Generation of magnetic fields}

The evolution of the Weibel instability can be understood as evolution
of current filaments generated in a plasma (Medvedev et al. 2005; Kato
2005).  The growth of the instability ceases once the magnitude of
current in each filament reaches the Alfv\'en current (Kato 2005);
physically this condition is consistent with the condition
by  Medvedev \& Loeb (1999)
that the particle's gyroradii in the excited magnetic fields are
comparable to the characteristic wavelength of the magnetic field. In
proton-electron plasmas, heavier protons mainly determine the final
magnetic field strength (Frederiksen et al. 2004). Thus, we consider the
generation of magnetic fields by protons. At a shock front, an
anisotropy is produced in the PDF of protons owing to the difference
between the shock speed $u$ and the thermal velocity of protons $v_{\rm
th}$. When the shock Mach number (${\cal M}=u/v_{\rm th}$) is larger
than $\sim 2$, which is the case in the galactic shocks we consider
later, the anisotropy becomes large enough to adopt the `particle-limit'
solution discussed by Kato (2005). In this case, the saturated magnetic
field is determined only by the shock velocity and the proton number
density $n_p$ as
\begin{equation}
\label{eq:mag} 
B_{\rm sat}\approx \chi_P u \sqrt{2\pi n_p m_p}\:,
\end{equation}
where $m_p$ is the proton mass, and $\chi_P$ is the isotropization
factor at the saturation, which is typically 0.5 (Kato 2005). The
saturated magnetic energy density reaches a sub-equipartition level with
the particle kinetic energy, which is consistent with previous
simulation results under large anisotropic conditions (Yang, Arons \&
Langdon 1994; Califano et al. 1998)

After saturation, the magnetic field strength decreases as the current
filaments merge. Recent particle simulations have shown that the final
magnetic field strength is given by $B_f=\eta_{\rm mer}^{1/2}B_{\rm
sat}$, where $\eta_{\rm mer}\sim 0.01$ (Silva et al. 2003; Medvedev et
al. 2005), thus we use this value. Long-term evolution (say Gyr) of the
generated magnetic fields is still an open question. However, Medvedev
et al. (2005) indicated that the fields do not decay rapidly after the
saturation, because the spatial scale of the fields increases rapidly
and dissipative processes such as Ohmic heating or ion-neutral
collisions, if any, are suppressed. The very long-term evolution of the
magnetic fields, for which the results of Medvedev et al. (2005) and
Silva et al. (2003) cannot be directly applied, is discussed in
Fujita and Kato (2005).

\subsection{Shock formation}

According to the standard hierarchical clustering scenario of the
universe, an initial density fluctuation of dark matter in the universe
gravitationally grows and collapses; its evolution can be approximated
by that of a spherical uniform over-dense region (Gunn \& Gott 1972;
Peebles 1980). The collapsed objects are called `dark halos' and the gas
in these objects later forms galaxies or clusters of galaxies. At the
collapse, the gas is heated by the `virial shocks' to the virial
temperature of the dark halo, $T_{\rm vir}=G M/(2 r_{\rm vir})$, where
$G$ is the gravitational constant, and $M$ and $r_{\rm vir}$ are the
mass and the virial radius of the dark halo, respectively. The relation
between the virial radius and the virial mass of an object is given by
\begin{equation}
 \label{eq:r_vir}
r_{\rm vir}=\left[\frac{3\: M}
{4\pi \Delta_c(z) \rho_{\rm crit}(z)}\right]^{1/3}\:,
\end{equation}
where $\rho_{\rm crit}(z)$ is the critical density of the universe and
$\Delta_c(z)$ is the ratio of the average density of the object to the
critical density at redshift $z$.  In the $\Lambda$CDM cosmology, the
critical density depends on redshift because the Hubble constant depends
on that, and it is given by
\begin{equation}
\label{eq:rho_crit}
 \rho_{\rm crit}(z)
=\frac{\rho_{\rm crit,0}\Omega_0 (1+z)^3}{\Omega(z)}\:,
\end{equation} 
where $\rho_{\rm crit,0}$ is the critical density at $z=0$, and
$\Omega(z)$ is the cosmological density parameter given by
\begin{equation}
 \Omega(z) = \frac{\Omega_0 (1+z)^3}{\Omega_0 (1+z)^3 + \lambda}
\end{equation}
for the flat universe with non-zero cosmological constant. The ratio
$\Delta_c(z)$ is given by
\begin{equation}
\label{eq:Dc_lam}
  \Delta_c(z)=18\:\pi^2+82 x-39 x^2\:, 
\end{equation}
for the flat universe (Bryan \& Norman 1998), where the parameter $x$ is
given by $x=\Omega(z)-1$. The virial shocks form at $r\approx r_{\rm
vir}$ and the velocity is $v_{\rm vir}\approx \sqrt{G M/r_{\rm vir}}$.

In addition, recent cosmological numerical simulations have shown that
`large-scale structure (LSS) shocks' form even before the gravitational
collapse (Miniati et al. 2000; Ryu et al. 2003; Furlanetto \& Loeb
2004); they form at the turnaround radius ($r_{\rm ta}\sim 2 r_{\rm
vir}$), the point at which the density fluctuation breaks off from the
cosmological expansion.  Thus, the gas that is going to form a galaxy or
a cluster undergoes two types of shocks; first, the gas passes the outer
LSS shock, and then, the inner virial shock. The typical velocity of the
LSS shocks is $v_s\approx H(z)R_p$ (Furlanetto \& Loeb 2004), where
$H(z)$ is the Hubble constant at redshift $z$, and $R_p$ is the physical
radius that the region would have had if it had expanded uniformly with
the cosmological expansion. The temperature of the postshock gas is
$T_s\approx 3/16\: (\mu m_p/k_B) v_s^2$, where $\mu m_p$ is the mean
particle mass, and $k_B$ is the Boltzmann constant. Note that we do not
consider mergers of objects that have already collapsed as the sites of
magnetic field generation.  During the merger, collapsed objects
just bring their magnetic fields to the newly born merged object.

Since the Weibel instability develops in ionized gas (plasma), we need
to consider the ionization history of the universe. After the entire
universe is ionized by stars and/or AGNs ($z\lesssim 8$), magnetic
fields are first generated at the LSS shocks. In this case, we do not
consider the subsequent generation of magnetic fields at the inner
virial shocks, because the strength is at most comparable to that of the
magnetic fields generated at the LSS shocks.  On the other hand, when
the universe is not ionized ($z\gtrsim 8$), the Weibel instability
cannot develop at the outer LSS shocks.  However, if the LSS shocks heat
the gas (mostly hydrogen) to $T_s>10^4{\rm\: K}$ and ionize it, the
instability can develop at the inner virial shocks.

For this case, the gas ionized at the LSS shock may recombine before it
reaches the virial shock. The recombination time-scale is given by
\begin{equation}
 \tau_{\rm rec} =\frac{1}{\alpha n_e}
\approx 1.22\times 10^5\; {\rm yr}\:\frac{1}{y}
\left(\frac{T}{\rm 10^4 K}\right)^{0.7}
\left(\frac{n_H}{\rm cm^{-3}}\right)^{-1}
\:,
\end{equation}
where $\alpha$ is the recombination coefficient, $n_e$ is the electron
density, $T$ is the gas temperature, $y$ is the ionization fraction, and
$n_H$ is the hydrogen density (Shapiro \& Kang 1987). If we assume that
$\tau_{\rm rec}=\tau_{\rm dyn}$, where $\tau_{\rm dyn}\approx
(1/2)r_{\rm ta}/u$ is the time-scale that the gas moves from the LSS
shock to the virial shock, the ionization rate when the gas reaches the
virial shock is
\begin{equation}
\label{eq:rec}
 y\approx \left(\frac{\tau_{\rm dyn}}
{1.22\times 10^5\; {\rm yr}}\right)^{-1}
\left(\frac{T}{\rm 10^4 K}\right)^{0.7}
\left(\frac{n_H}{\rm cm^{-3}}\right)^{-1}
\;
\end{equation}
for $y<1$. We found that $y<1$ for $z\gtrsim 9$, and the minimum value
when the generation of magnetic fields is effective ($z\lesssim 12$, see
\S\ref{sec:results}) is $y\sim 0.3$. When $y<1$, we simply replace $n_p$
in equation~(\ref{eq:mag}) with $y n_H$. For temperature, we assumed
$T=T_s$ in equation~(\ref{eq:rec}). Since the magnetic fields do not
much depend on the temperature ($B_{\rm sat}\propto y^{0.5}\propto
T^{0.35}$), they do not much change even when radiative cooling reduces
the temperature; at least $B_{\rm sat}$ does not change by many orders
of magnitude. Since the ionization rate is fairly large, the ambipolar
diffusion of magnetic fields can be ignored (e.g. eq.~13--57 in Spitzer
1978). It was shown that the ionization rate just behind a shock is
$y\sim 0.1$, if the shock velocity is relatively small $40\rm\; km\;
s^{-1}$ (Shapiro \& Kang 1987; Susa et al. 1998) . If the shock
velocity is larger, $y\sim 1$. In our calculations, the velocity of the
LSS shocks is $u\gtrsim 40\rm\; km\; s^{-1}$, when the generation of
magnetic fields is effective. Thus, $y$ just behind the shocks is at
least comparable to that obtained through the condition of $\tau_{\rm
rec}\sim \tau_{\rm dyn}$, and the incomplete ionization does not much
affect the results shown in the next section.

\section{Results and discussion}
\label{sec:results}

\begin{figure}
\resizebox{\hsize}{!}
{\includegraphics[]{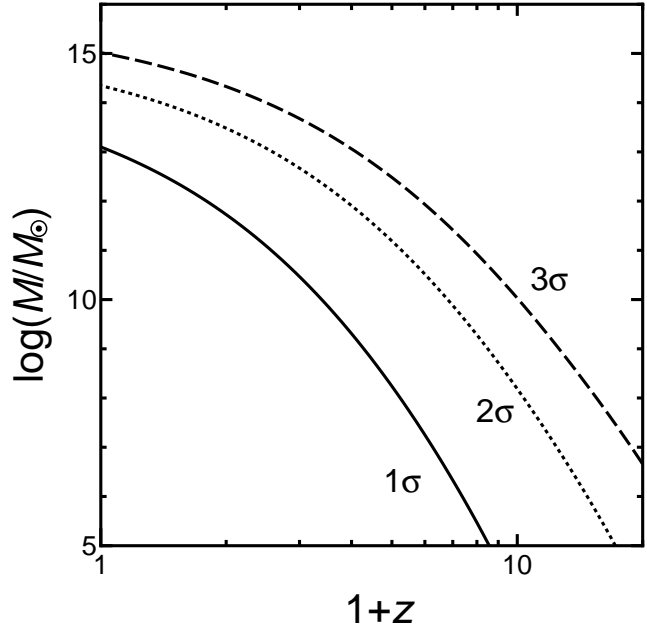}} \caption{Typical masses of objects
forming at redshift $z$. The labels, $1\sigma$, $2\sigma$, and $3\sigma$
(solid, dotted, and dashed lines, respectively) indicate the amplitudes
of initial density fluctuations from which the objects formed; $\sim
1$--$3\sigma$ is the typical value (Barkana \& Loeb 2001). Objects with
masses of $M\lesssim 10^{12}\:M_\odot$ and 
$M\gtrsim 10^{13}\:M_\odot$ correspond
to galaxies (ellipticals and spirals) and clusters of galaxies,
respectively. } 
\label{fig:mass}
\end{figure}

\begin{figure}
\resizebox{\hsize}{!}  {\includegraphics[]{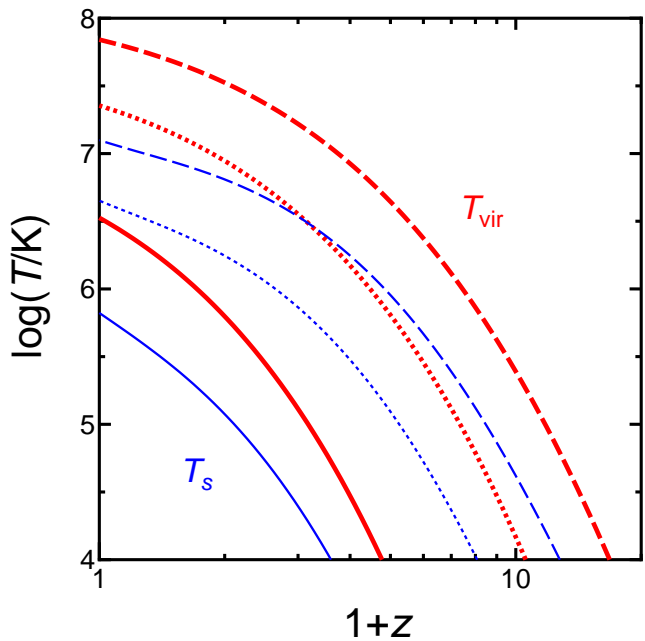}}
\caption{Temperatures behind the virial shocks ($T_{\rm vir}$; thick
lines) and those behind the LSS shocks ($T_s$; thin lines) for objects
forming at redshift $z$ (Fujita \& Kato 2005). Solid, dotted, and dashed
lines correspond to $1\sigma$, $2\sigma$, and $3\sigma$ fluctuations,
respectively (see Fig.\ref{fig:mass})} \label{fig:temp}
\end{figure}

\begin{figure}
\resizebox{\hsize}{!}  {\includegraphics[]{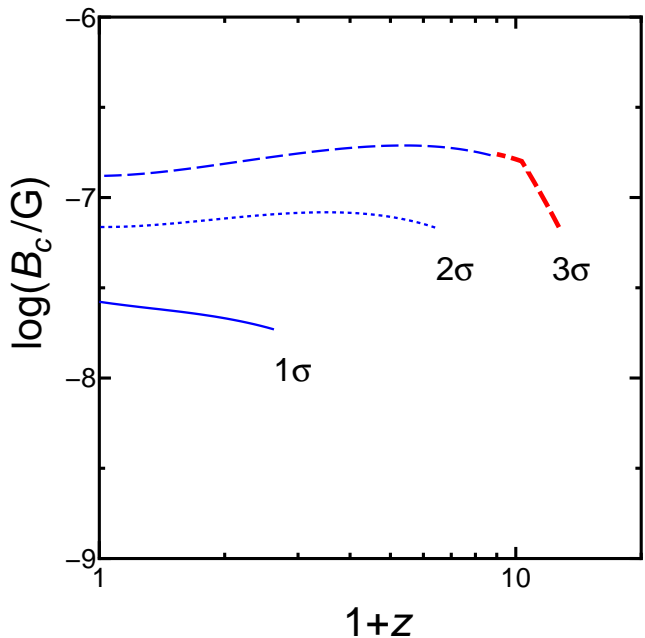}} \caption{Typical
magnetic field strengths of objects forming at redshift $z$ (Fujita \&
Kato 2005). Solid, dotted, and dashed lines correspond to $1\sigma$,
$2\sigma$, and $3\sigma$ fluctuations, respectively (see
Fig.\ref{fig:mass}).}  \label{fig:B}
\end{figure}

Fig.\ref{fig:mass} shows the typical mass of objects, $M$, as a function
of redshift $z$; the labels $1\sigma$, $2\sigma$, and $3\sigma$ indicate
the amplitudes of initial density fluctuations in the universe from
which the objects form, on the assumption of the CDM fluctuations
spectrum (Barkana \& Loeb 2001). Fig.\ref{fig:temp} shows the downstream
temperature at the virial shock ($T_{\rm vir}$) and that at the LSS
shock ($T_s$) for the objects; $T_{\rm vir}$ is always larger than
$T_s$. The ratio $T_{\rm vir}/T_s$ indicates that ${\cal M}\gtrsim 4$
for the virial shock. For the LSS shock, ${\cal M}\gg 1$, because the
gas outside the shock is cold. Thus, equation~(\ref{eq:mag}) can be
applied to both shocks. In Fig.\ref{fig:B}, we present the strength of
magnetic fields ($B_c$) at a scale of $r_{\rm vir}$ for the collapsed
objects. We assume that the entire universe is reionized at $z=8$. Thus,
for $z>8$, magnetic fields are generated only at the virial shocks if
$T_s>10^4\rm\: K$. We assume that $B_c=B_f$ and plot the lines only when
$T_s>10^4\rm\: K$. The recombination is effective at $z>12$ for the
$3\sigma$ model and at $z>8$ for the $2\sigma$ model. On the other hand,
for $z<8$, the magnetic fields are generated at the LSS shocks. We
consider the compression of the fields while the size of the gas sphere
decreases from $r=r_{\rm ta}$ to $r=r_{\rm vir}$, and thus we assume
that $B_c=8^{2/3}\: B_f$. Moreover, we plot the lines only for $T_{\rm
vir}>2\times 10^5$~K, because below this temperature, gas infall is
suppressed by photoionization heating (Efstathiou 1992; Furlanetto \&
Loeb 2004).

In Fig.\ref{fig:B}, the strength of magnetic fields reaches $\sim
10^{-8}$--$10^{-7}$~G, which is very close to the values observed in
nearby galaxies and clusters of galaxies ($\sim 10^{-6}$~G; Clarke et
al. 2001; Widrow 2002; Vall{\' e}e 2004). This indicates that strong
amplification of magnetic fields, such as dynamo amplification, is not
required after formation of the galaxies and clusters. The absence of a
strong amplification is consistent with the observations of galactic
magnetic fields at $z\gtrsim 2$ (Athreya et al. 1998). Future
observations of higher-redshift galaxies would discriminate between our
model and strong dynamo amplification models; the latter predict much
weaker magnetic fields at higher redshifts. Moreover, since the
predicted galactic magnetic fields are comparable to those at present,
they might have affected the formation of stars in
protogalaxies. Fig.\ref{fig:B} also shows that our model naturally
explains the observational fact that the magnetic field strengths of
galaxies and galaxy clusters fall in a small range (a factor of
10). Since our model predicts that magnetic fields are generated around
objects, magnetic fields in intergalactic space are not required as the
seed or origin of galactic magnetic fields.

Although it would be difficult to directly observe the generation of
magnetic fields through the Weibel instability for distant high-redshift
galaxies, it would be easier for nearby clusters of galaxies. Since
clusters are now in the formation process, LSS shocks should be
developing outside of the virial radii of the clusters (Miniati et
al. 2000; Ryu et al. 2003). Since particles are often accelerated at
shocks, the synchrotron emission from the accelerated particles could be
observed with radio telescopes with high sensitivity at low frequencies
(Keshet, Waxman \& Loeb 2004). The total non-thermal luminosity due to
synchrotron luminosity plus inverse Compton scattering with cosmic
microwave background (CMB) photons is estimated as
\begin{equation}
 L_{\rm nt}\approx 
\frac{\epsilon}{r_{\rm ta}/u} \frac{f M}{m_p}\frac{1}{2}u^2\:,
\end{equation}
where $\epsilon$ is the acceleration efficiency and $f$ is the gas
fraction of a cluster. If we assume $\epsilon=0.03$ and $f=0.15$, the
maximum value of $L_{\rm nt}$ is $\sim 10^{43}\rm\: erg\: s^{-1}$. Since
the energy density of magnetic fields ($u_B$) is smaller than that of
the CMB ($u_{\rm CMB}$), most of the non-thermal luminosity ($L_{\rm
nt}$) is attributed to the inverse Compton scattering such as $u_{\rm
CMB}L_{\rm nt}/(u_{\rm B}+u_{\rm CMB})$, which may have been detected in
the hard X-ray band ($\gtrsim 20$~keV; Fusco-Femiano et al. 2004). Thus,
the synchrotron radio luminosity is $u_B L_{\rm nt}/(u_{\rm B}+u_{\rm
CMB})\lesssim 10^{41}\rm\: erg\: s^{-1}$. Some of the diffuse radio
sources observed in the peripheral cluster regions (`radio relics') may
be this emission (Govoni \& Feretti 2004). Since the Weibel instability
generates magnetic fields on the plane of the shock front, the
synchrotron emission should be polarized perpendicular to the shock
front (Medvedev \& Loeb 1999), which is actually observed for some radio
relics (Govoni \& Feretti 2004). The synchrotron emission will tell us
the positions of the LSS shocks, if they exist.  If magnetic fields are
generated there, they should be observed only downstream of the
shock. This may be confirmed through Faraday rotation measurements of
radio sources behind the cluster for both sides of the shock, if the
coherent length of the fields sufficiently increases.

\section{Japanese X-ray missions}

\subsection{Suzaku (Astro-E2)}

\begin{figure}
\resizebox{\hsize}{!}  {\includegraphics[]{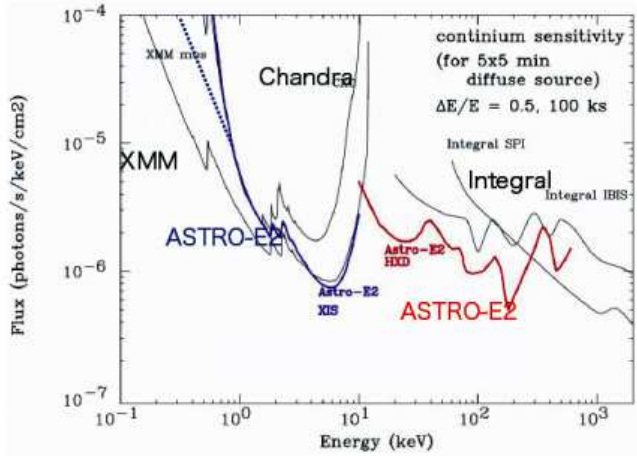}} \caption{The
sensitivities of {\it Suzaku} ({\it Astro-E2}) and other satellites.}
\label{fig:suzaku_sen}
\end{figure}

{\it Suzaku} is the recovery mission for {\it ASTRO-E}, which did not
achieve orbit during the launch in 2000. {\it Suzaku} was launched in
July 10, 2005. {\it Suzaku} has three detectors ({\it XRS}, {\it XIS},
and {\it HXD}). {\it XRS} has a superb energy resolution 
($\lesssim$ 6.5 eV $\sim 
100\rm\: km\: s^{-1}$ ). We expected that it could detect
gas motion in clusters, for example. {\it XRS} uses liquid helium to
cool the detector.  Unfortunately, the helium evaporated before the
first light (Aug.8) and {\it XRS} was lost. The cause of the trouble is
now under investigation. The observation schedule will be changed and
optimized to the remaining two detectors ({\it XIS} and {\it HXD}).
{\it XIS} has imaging capability and its energy range is
0.2--12~keV. {\it HXD} is a hard X-ray detector and its energy range is
10--600~keV. {\it XIS} and {\it HXD} are working well. Their sensitives
are shown in Figure~\ref{fig:suzaku_sen}. It is to be noted that the
background levels of these detectors are lower than those of {\it
Chandra} and {\it XMM-Newton}.  {\it XIS} and {\it HXD} will give us
various information in the field of high energy astrophysics (e.g. hard
X-ray from black holes and clusters of galaxies). Readers can get the
information on {\it Suzaku} at several web
sites.\footnote{http://www.isas.jaxa.jp/e/enterp/missions/astro-e2/,
http://heasarc.gsfc.nasa.gov/docs/astroe/astroe2.html}

\subsection{NeXT}

\begin{figure}
\resizebox{\hsize}{!}  {\includegraphics[]{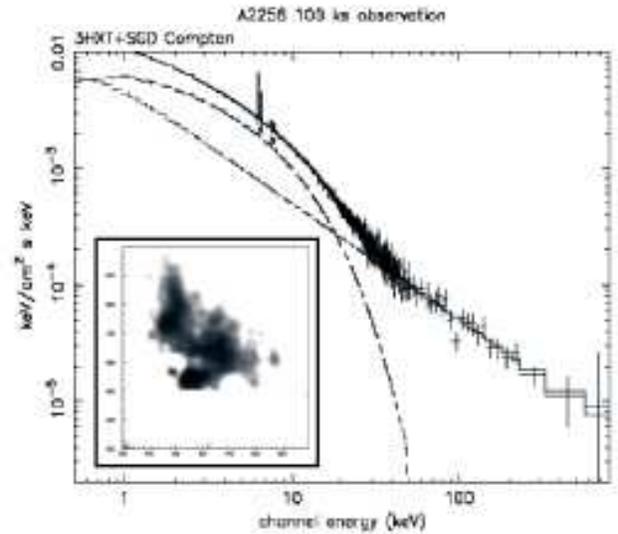}} \caption{
 A simulated hard
 X-ray image and the spectrum of the radio relic in Abell~2256 observed by
 {\it NeXT}.}
\label{fig:a2256}
\end{figure}

\begin{figure}
\resizebox{\hsize}{!}  {\includegraphics[]{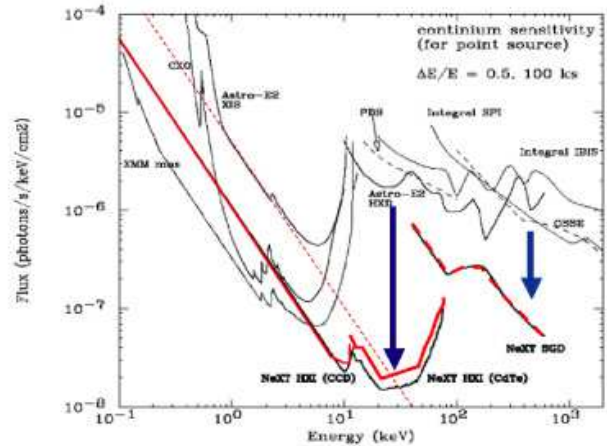}} \caption{The
aimed sensitivity of {\it NeXT} and the sensitivities of other
 satellites. }
\label{fig:next_sen}
\end{figure}

{\it NeXT} ({\it New X-ray Telescope }) is the major Japanese X-ray
mission after {\it
Suzaku}.\footnote{http://www.astro.isas.ac.jp/future/NeXT/doc/concept.html}
If the plan is approved, {\it NeXT} will be launched in 2011, before the
launch of {\it Constellation-X} and {\it XEUS}. {\it NeXT} will have
imaging capability in the hard X-ray band. For example, we will easily
get the images of radio relics in clusters of galaxies by observing
their inverse Compton emission (Figure~\ref{fig:a2256}). The sensitivity
will be much higher than those of previous satellites
(Figure~\ref{fig:next_sen}). {\it NeXT} will also have an ultra-high
resolution soft X-ray spectrometer, which will be another challenge of
{\it Suzaku XRS}. Moreover, {\it NeXT} will have a soft gamma-ray
detector. Thus, it will enable us to observe objects in a very wide
energy band.

\acknowledgements

We thank K.~Omukai, N.~Okabe, T.~Kudoh, K.~Asano, S.~Inoue, K.~Nakazawa,
and H. Matsumoto for discussions. Y. F. is supported in part by a
Grant-in-Aid from the Ministry of Education, Culture, Sports, Science,
and Technology of Japan (14740175).


\end{document}